\documentstyle[preprint,pre,aps]{revtex}

\newcounter{saveeqn}

\title{Image restoration using the Q-Ising spin glass}

\author{Jun-ichi Inoue} 
\address{Complex Systems Engineering, Graduate School of 
Engineering, \\
Hokkaido University, N13-W8, Kita-ku, Sapporo, 060-8628, Japan}
\author{Domenico M. Carlucci} 
\address{Instituut voor Theoretische Fysica,\\
K. U. Leuven, B-3001 Leuven, Belgium }

\date{\today}
\begin{document}
\maketitle
\thispagestyle{empty}
\begin{abstract}
We investigate static and dynamic properties of gray-scale 
image restoration (GSIR) by making use of the Q-Ising spin glass model, 
whose ladder symmetry allows to take in account the distance between two spins.
We thus give an explicit expression of the Hamming distance between 
the original and restored images as a function of the hyper-parameters in the 
mean field limit.  Finally, numerical simulations for real-world pictures  
are carried out to prove the efficiency of our model.   
\end{abstract}
\mbox{}
\clearpage
\setcounter{page}{1} 
\section{Introduction}
In the last decade, the problem of the image restoration (IR) has 
been successfully investigated by means of techniques borrowed from the field of 
statistical mechanics. Among them, 
it is certainly worthy mentioning 
the {\it Maximum Posterior 
Marginal} (MPM) and {\it Maximum A Posteriori} (MAP) 
estimations.
From the statistical mechanical point of view, each recovered image 
within the MPM estimation
 can be regarded as the equilibrium 
state of ferromagnetic spin systems in 
the presence of random fields at finite temperature. 
In simple words, the reconstruction of a corrupted image is achieved 
by balancing the strength of a linear field, which  carries the information
of the degraded picture, and a ferromagnetic term which 
builds relatively large ``one-color'' clusters 
(below the transition temperature), thus suppressing 
the isolated pixels thought to be noise. From this point of view, 
the MAP estimation consists in the minimization of the same Hamiltonian  
at zero temperature (search for the ground state), with an appropriate 
scaling of the random field. 
The advantage of the MPM estimation over the MAP one 
has been  pointed out 
by Marroquin et al \cite{Marroquin} and its 
performance has been investigated by several authors\cite{PB,GG}. 
In this direction,  Nishimori and Wong\cite{NW},  by unifying  
IR problem and error-correcting code theory under a single framework, 
found that the optimal recovering of an image is obtained 
at a finite temperature (known as {\it Nishimori temperature} in the field of 
statistical mechanics).
Their results, however,  were restricted to the usual binary spin models 
(Ising), {\it i.e.}  black or white images in IR jargon,  
and many questions about the properties of the gray-scale image 
restoration  (GSIR) processes still remain open. 
A first attempt to generalize \cite{NW} to gray-level pictures 
has been carried out by the authors  
\cite{CI} by mapping the set of the pixels 
onto Q-state (chiral) Potts spins 
in the presence of the random fields. 
In that case, the symmetry of the Potts Hamiltonian ({\it hyper-tetrahedron}) 
reduces the problem to a 2-state-{\it like} system, where only one 
bit turns out to be right, and all the others are equivalently 
wrong without any regard of the whole gray-level scale. 
Whereas this turns out to be an efficient method in the presence of white 
noise (each spin is flipped to any of the Q values with equal probability),  
things may be  different from a transmission channel affected by 
Gaussian noise (the spin-flip probability distribution is a Gaussian).
We thus investigate the performance (both static and dynamic) 
of the gray-scaled image restoration using the Q-Ising model \cite{Bolle}, 
whose {\it ladder} symmetry takes in account the distance between 
the spin values and will allow us to say, for instance,  that $Q=3$ 
is better than $Q=5$ if the right pixel corresponds to $Q=2$.
The analytical expressions are obtained in so-called mean field 
limit, where each spin interacts with all the others. 
The efficiency of our model is 
checked by Monte Carlo simulations and 
iterative algorithm by using mean-field 
approximation.
This paper is organized as follow. 
In the next section, we introduce the 
infinite range Q-Ising model and
in Sec. \ref{Statics} we give an analytical 
expression of the Hamming 
distance in the mean field limit. 
In Sec. \ref{Dynamics}  we derive the dynamical equations with 
respect to the macroscopic quantities, namely, 
the magnetization and the Hamming distance 
in terms of microscopic master equation. 
In Sec. \ref{MC}, in order to test the usefulness of 
the Q-Ising model for the GSIR, we carry out Monte Carlo simulations 
for real-world pictures with $Q=8$ gray-scale levels.
In  Sec.  \ref{It}, we show an iterative algorithm based on the mean-field 
approximation, whose convergence is much faster than that of the 
Monte Carlo simulations. 
The last section is left for summary and discussions.
%
\section{The infinite range Q-Ising spin glass model \label{infinite_range}}
%
A Q-gray scale levels image is nothing else that a set of pixels 
$\{\xi \}$ on 
a grid, whose values can be coded at each node as 
an integer variable $\xi_{i}
\in \{1,2,{\cdots},Q\}$. Without loss of generality,  
let our image be generated by the following prior distribution 
\begin{eqnarray}
P_{s}(\{ \xi \}) & = & 
\frac{1}{{\cal Z}_{s}}
\,{\exp}\left[
-\frac{\beta_{s}}{2}
\sum_{ij}(\xi_{i}-\xi_{j})^{2}
\right], 
\label{prior}
\end{eqnarray}
where ${\cal Z}_{s}$ is the usual normalization constant 
that is given by 
\begin{eqnarray}
{\cal Z}_{s} & = & 
{\rm tr}_{\{\xi\}}
{\exp}\left[
-\frac{\beta_{s}}{2}
\sum_{ij}(\xi_{i}-\xi_{j})^{2}
\right].
\end{eqnarray}
In the spirit of statistical mechanics, we want to  regard this picture as  
a snapshot of a spin system described by the Hamiltonian 
${\cal H}_{s}\,\equiv\,
(1/2)\sum_{ij}(\xi_{i}-\xi_{j})^{2}$ 
at a specific temperature $T_{s}\,\equiv\, 
{\beta}_{s}^{-1}$.
Sending our image through a noisy channel  will cause the flipping 
of some pixels to different values.
For this degrading process, 
we assume that each pixel $\xi_{i}$ changes 
its state to $\tau_{i}$ independently. 
Then, the degraded pixel $\tau_{i}$ 
is given by the following conditional 
probability 
\begin{eqnarray}
P(\tau_i|\xi_i)=  
\frac{1}{\sqrt{2\pi}\tau}
\,{\exp}
\left[
-\frac{1}{2\tau^{2}}
(\tau_{i}-\tau_{0}\xi_{i})^{2}
\right].
\end{eqnarray}
This means that after the transmission, 
the receiver observes $\tau_{i}$ which was 
violated from scaled original 
image $\tau_{0}\xi_{i}$ with 
a standard deviation $\tau$. 
This kind of damaging process is referred to as 
``Gaussian channel (GC)''.
Due to the independence 
of noisy process on each pixel, a sequence of 
original pixel $\{\xi\}$ is corrupted by the GC as 
\begin{eqnarray}
P(\{\tau\}|\{\xi\}) & = & \prod_{i}P(\tau_{i}|\xi_{i}) \nonumber \\
\mbox{} & = & 
\frac{1}{\sqrt{2\pi}\tau}
{\exp}\left[
-\frac{1}{2\tau^2}
\sum_{i}(\tau_{i}-\tau_{0}\xi_{i})^{2}
\right].
\end{eqnarray}
In the context of Bayesian approach, 
the probability that the estimate of the 
source sequence is $\{ \sigma \}$ 
provided that the observed noisy data is 
$\{ \tau \}$, 
reads 
\begin{eqnarray}
P(\{ \sigma \}|\{ \tau \}) & = & 
\frac{P(\{ \tau \}|\{ \sigma \})
P(\{ \sigma \})}
{{\rm tr}_{\{ \sigma \}}
P(\{ \tau \}|\{ \sigma \})
P(\{ \sigma \})} \nonumber \\
\mbox{} & = & 
\frac{{\exp}\left[
-h\sum_{i}(\sigma_{i}-\tau_{i})^{2}
-(\beta_{d}/2)\sum_{ij}(\sigma_{i}-\sigma_{j})^{2}
\right]}
{{\rm tr}_{\{\sigma\}}{\exp}\left[
-h\sum_{i}(\sigma_{i}-\tau_{i})^{2}
-(\beta_{d}/2)\sum_{ij}(\sigma_{i}-\sigma_{j})^{2}
\right]} \nonumber \\
\mbox{} & {\equiv} & \frac{{\exp}
\left(-{\cal H}_{\rm eff}
\right)}
{{\cal Z}_{d}}
\label{posterior1}
\end{eqnarray}
where we defined the effective Hamiltonian 
${\cal H}_{\rm eff}$ 
and the normalization constant 
${\cal Z}_{d}$ as 
\begin{eqnarray}
{\cal H}_{\rm eff} & \equiv & 
h\sum_{i}(\sigma_{i}-\tau_{i})^{2}
+\frac{\beta_{d}}{2}\sum_{ij}(\sigma_{i}-\sigma_{j})^{2}
\label{effHam0}
\end{eqnarray}
and
\begin{eqnarray}
Z_{d} & \equiv & 
{\rm tr}_{\{\sigma\}}
{\exp}\left[
-{\cal H}_{\rm eff}
\right], 
\end{eqnarray}
respectively.
The parameters 
$h$ and $\beta_{d}$ appearing 
in the Hamiltonian ${\cal H}_{\rm eff}$ 
[Eq. (\ref{effHam0})] 
are referred to as 
``hyper-parameters'' and we can not 
mention about the true values of them 
beforehand. 
This conditional probability 
$P(\{\sigma\}|\{\tau\})$ is 
called 
``posterior probability'' and is 
constructed in terms of 
a likelihood $P(\{\tau\}|\{\sigma\})$ and 
a prior probability $P(\{\sigma\})$ as 
we saw in Eq. (\ref{posterior1}).  
$P(\{\tau\}|\{\sigma\})$ and 
$P(\{\sigma\})$ are given by 
\begin{eqnarray}
P(\{\tau\}|\{\sigma\}) & = & 
\frac{{\exp}\left[
-h\sum_{i}(\tau_{i}-\sigma_{i})^{2}
\right]}
{{\rm tr}_{\{\tau\}}{\exp}\left[
-h\sum_{i}(\tau_{i}-\sigma_{i})^{2}
\right]}
\end{eqnarray}
and
\begin{eqnarray}
P(\{\sigma\}) & = & 
\frac{{\exp}\left[
-(\beta_{d}/2)\sum_{ij}(\sigma_{i}-\sigma_{j})^{2}
\right]}
{{\rm tr}_{\{\sigma\}}{\exp}\left[
-(\beta_{d}/2)\sum_{ij}(\sigma_{i}-\sigma_{j})^{2}
\right]}.
\end{eqnarray}
The prior probability 
reflects our assumption on the original image 
that the picture should be locally smooth. 
As shortly mentioned in the introduction, the MAP 
estimation consists in maximizing the 
above posterior probability $P(\{ \sigma \}|\{ \tau \})$, that is  
finding the ground state $\{ \sigma \}$ of 
the effective Hamiltonian ${\cal H}_{\rm eff}$ and 
regarding it as an estimate of true pixels. 

On the other hand, 
in the context of the MPM estimation, 
we first consider the following marginal distribution; 
\begin{eqnarray}
{P}({\sigma}_{i}|\{ \sigma \})=
\sum_{\sigma \neq \sigma_{i}}
P(\{ \sigma \}|\{ \tau \})
\end{eqnarray}
and then we calculate the local magnetization which is given by 
\begin{eqnarray}
\langle {\sigma}_{i} \rangle_{\beta_{d},h} & \equiv  & 
\sum_{\sigma_{i}=1}^{Q}{\sigma}_{i}{P}(\sigma_{i}|
\{ \tau \}) \nonumber \\
\mbox{} & = & \frac{{\rm tr}_{\{\sigma\}}
\sigma_{i} \sum_{\sigma \neq \sigma_{i}} 
{\exp}\left[
-h\sum_{i}(\tau_{i}-\sigma_{i})^{2}
-(\beta_{d}/2)\sum_{ij}(\sigma_{i}-\sigma_{j})^{2}
\right]}
{{\rm tr}_{\{\sigma\}}
{\exp}\left[
-h\sum_{i}(\tau_{i}-\sigma_{i})^{2}
-(\beta_{d}/2)\sum_{ij}(\sigma_{i}-\sigma_{j})^{2}
\right]} \nonumber \\
\mbox{} & = & 
\frac{{\rm tr}_{\{\sigma\}}
\sigma_{i}{\exp}\left[
-h\sum_{i}(\tau_{i}-\sigma_{i})^{2}
-(\beta_{d}/2)\sum_{ij}(\sigma_{i}-\sigma_{j})^{2}
\right]}
{{\rm tr}_{\{\sigma\}}
{\exp}\left[
-h\sum_{i}(\tau_{i}-\sigma_{i})^{2}
-(\beta_{d}/2)\sum_{ij}(\sigma_{i}-\sigma_{j})^{2}
\right]} \nonumber \\
\mbox{} & = & 
\frac{{\rm tr}_{\{\sigma\}}
\sigma_{i}{\exp}\left[
-{\cal H}_{\rm eff}
\right]}
{{\rm tr}_{\{\sigma\}}
{\exp}\left[
-{\cal H}_{\rm eff}
\right]}.
\label{local_m1}
\end{eqnarray}
Using the above expectation value, we 
regard the estimate of the original pixel $\xi_{i}$ 
as ${\Omega}(\langle 
\sigma_{i}\rangle_{\beta_{d},h})$ where function 
${\Omega}$ is represented by 
a sum of step functions $\Theta (x)$ ($\Theta(x)=1$ for $x \geq 0$ 
and $\Theta(x)=0$ for $x<0$);  
\begin{eqnarray}
{\Omega}(\langle \sigma_{i} \rangle_{\beta_{d},h})  & {\equiv} & 
\sum_{k=1}^{Q}
k{\biggr [}
{\Theta}
\left(
\langle \sigma_{i} \rangle_{\beta_{d},h} -\frac{2k-1}{2}
\right) - {\Theta}\left(
\langle \sigma_{i} \rangle_{\beta_{d},h}-\frac{2k+1}{2}
\right)
{\biggr ]}.
\end{eqnarray}
The natural quantity measuring the quality of our restoration process, 
{\it viz.} the distance between the original and the recovered image, 
is the Hamming distance (square error)  
\begin{eqnarray}
D_{\rm H}(\beta_{d},h) \equiv   
\frac{1}{2N}\sum_{i}
[\xi_{i}-{\Omega}(\langle \sigma_{i} \rangle)_{\beta_{d},h}]^{2}  
\end{eqnarray}
whose value depends upon the hyper-parameters, $h,\beta_{d}$ 
appearing in the effective Hamiltonian ${\cal H}_{\rm eff}$. 
At this stage, it is important  to bear in mind that 
the MAP estimate is recovered as the 
limit  $\beta_{d}{\rightarrow} \infty$ (keeping their ratio constant 
$H \equiv h/\beta_{d}$) in Eq. (\ref{local_m1}).
Encouraged by the results in \cite{NW} and \cite{CI}, 
we expect that more data fed through the noisy channel 
improves the quality of the restored image, since the 
receiver will have more information about the original image.
Therefore, in addition to the transmission of the single bit 
$\xi_i$, we send also 
the pairwise product $\xi_i\xi_j$.
Then, 
each product $\xi_{i}\xi_{j}$ is also 
corrupted independently by the following GC 
\begin{eqnarray}
P(J_{ij}|\xi_{i}\xi_{j}) & = & 
\frac{1}{\sqrt{2\pi}J}
{\exp}\left[
-\frac{1}{2J^{2}}
(J_{ij}-J_{0}\xi_{i}\xi_{j})^{2}
\right], 
\end{eqnarray}
namely, the degraded version of the product $J_{ij}$  
deviated from the scaled original data $J_{0}\xi_{i}\xi_{j}$ 
with width $J$. 
For this degrading process, we modify a likelihood 
$P(\{\tau\}|\{\sigma\})$ in Eq. (\ref{posterior1}) as 
\begin{eqnarray}
P(\{\tau\},\{J\}|\{\sigma\}) & = & 
\frac{{\exp}\left[
-(\beta_{J}/2)\sum_{ij}(J_{ij}-\sigma_{i}\sigma_{j})^{2}
-h\sum_{i}(\tau_{i}-\sigma_{i})^{2}
\right]}
{{\cal Z}_{L}^{'}}
\end{eqnarray}
with 
\begin{eqnarray}
{\cal Z}_{L}^{'} & \equiv & 
{\rm tr}_{\{\tau\},\{J\}}
{\exp}\left[
-\frac{\beta_{J}}{2}
\sum_{ij}(J_{ij}-\sigma_{i}\sigma_{j})^{2}
-h\sum_{i}(\tau_{i}-\sigma_{i})^{2}
\right]
\end{eqnarray}
where 
we introduced another hyper-parameter $\beta_{J}$.  
Using the same way as Eq. (\ref{posterior1}), 
we rewrite the posterior probability as 
\begin{eqnarray}
P(\{\sigma\}|\{\tau\},\{J\}) & = & 
\frac{{\exp}\left[-{\cal H}_{\rm eff}\right]}
{{\rm tr}_{\{\sigma\}}
{\exp}\left[-{\cal H}_{\rm eff}\right]}
\end{eqnarray}
with the following effective Hamiltonian 
\begin{eqnarray}
{\cal H}_{\rm eff} & = & 
\frac{\beta_{J}}{2}
\sum_{ij}(J_{ij}-\sigma_{i}\sigma_{j})^{2}
-h\sum_{i}(\tau_{i}-\sigma_{i})^{2}-
\frac{\beta_{d}}{2}
\sum_{ij}(\sigma_{i}-\sigma_{j})^{2}.
\label{effectHam1}
\end{eqnarray}
Given the degraded version 
of data, namely, $\{\tau\}$ and $\{J\}$, 
arbitrary macroscopic physical quantity 
$f(\{\sigma\},\{\tau\},\{J\})$ 
is calculated in terms of the average over 
the posterior distribution 
$P(\{\sigma\}|\{\tau\},\{J\})$ as 
\begin{eqnarray}
\langle f(\{\sigma\},\{\tau\},\{J\}) \rangle_{\beta_{d},h} & \equiv & 
{\rm tr}_{\{\sigma\}}
f(\{\sigma\},\{\tau\},\{J\})
P(\{\sigma\}|\{\tau\},\{J\}) \nonumber \\
\mbox{} & = & 
\frac{{\rm tr}_{\{\sigma\}}
f(\{\sigma\},\{\tau\},\{J\})
{\rm e}^{-{\cal H}_{\rm eff}}}
{{\rm tr}_{\{\sigma\}}
{\rm e}^{-{\cal H}_{\rm eff}}}.
\label{f_ave1}
\end{eqnarray}
As the quantity 
$\langle f(\{\sigma\},\{\tau\},\{J\}) \rangle_{\beta_{d},h}$ 
depends on the observed data 
$\{\tau\},\{J\}$, we should average them out by 
the distribution 
\begin{eqnarray}
\mbox{} & \mbox{} & P(\{\tau\},\{J\}|\{\xi\}) \nonumber \\
\mbox{} & = & 
\frac{{\exp}\left[
-\frac{1}{2J^{2}}
\sum_{ij}(J_{ij}-J_{0}\xi_{i}\xi_{j})^{2}
-(1/2\tau^{2})\sum_{i}(\tau_{i}-\xi_{i})^{2}
-\frac{\beta_{s}}{2}
\sum_{ij}(\xi_{i}-\xi_{j})^{2}
\right]}
{{\rm tr}_{\{\tau\},\{J\},\{\xi\}}{\exp}\left[
-\frac{1}{2J^{2}}
\sum_{ij}(J_{ij}-J_{0}\xi_{i}\xi_{j})^{2}
-(1/2\tau^{2})\sum_{i}(\tau_{i}-\xi_{i})^{2}
-\frac{\beta_{s}}{2}
\sum_{ij}(\xi_{i}-\xi_{j})^{2}
\right]}.
\end{eqnarray}
Therefore, after tracing the original image $\{\xi\}$ out, 
the averaged macroscopic quantity is given by 
\begin{eqnarray}
\mbox{} & \mbox{} & [\langle f(\{\sigma\},\{\tau\},\{J\}) 
\rangle_{\beta_{d},h}]_{\{\tau\},\{J \},\{\xi\}} \nonumber \\
\mbox{} & \equiv & 
{\rm tr}_{\{\tau\},\{J\},\{\xi\}}
\left[
\frac{{\rm tr}_{\{\sigma\}}
f(\{\sigma\},\{\tau\},\{J\})
{\rm e}^{-{\cal H}_{\rm eff}}}
{{\rm tr}_{\{\sigma\}}
{\rm e}^{-{\cal H}_{\rm eff}}}
\right]
P(\{\tau\},\{J\}|\{\xi\})
\label{f_ave2}
\end{eqnarray}
Using this definition, 
the performance 
of image restoration is measured by the following 
averaged Hamming distance between 
the original image and the restored one, that is,   
$\Omega(\langle \sigma_{i} \rangle_{\beta_{d},h})$ as 
\begin{eqnarray}
D_{\rm H} & \equiv & 
{\rm tr}_{\{\tau\},\{J\},\{\xi\}}
\left[
\frac{{\rm tr}_{\{\sigma\}}
(1/2N)\sum_{i}(\xi_{i}-
\Omega(\langle \sigma_{i} \rangle_{\beta_{d},h})
)^{2}{\rm e}^{-{\cal H}_{\rm eff}}}
{{\rm tr}_{\{\sigma\}}
{\rm e}^{-{\cal H}_{\rm eff}}}
\right]P(\{\tau\},\{J\}|\{\xi\}). 
\label{aveHam1}
\end{eqnarray}
In the next two subsection, 
we investigate the performance of image restoration 
in terms of this Hamming distance $D_{H}$.
We focus our analysis not 
only on the static properties 
but also the dynamic properties 
of image restoration. 
\subsection{Static properties \label{Statics}}

In this subsection, we consider the static properties 
of image restoration. 
First of all, we should investigate 
the properties of original image, that is to say, 
the properties of the ferro magnetic Q-Ising model. 
However, it is quite hard to calculate 
the partition function or the other 
physical quantities for our spin system defined 
on two dimensional square lattice analytically. 
Therefore, in this paper, we investigate the 
infinite range version of our model system and 
calculate the macroscopic physical quantities 
analytically. 
Then, the infinite range version of the 
prior distribution leads to 
\begin{eqnarray}
P_{s}(\{\xi\}) & = & 
\frac{1}{{\cal Z}_{s}}
{\exp}\left[
-\frac{\beta_{s}}{2N}
\sum_{ij}(\xi_{i}-\xi_{j})^{2}
\right],
\end{eqnarray}
where we should notice that the argument of the 
exponential should be divided by $N$ in order to 
take a proper thermodynamic limit. 
For this rather artificial model, 
we easily obtain the magnetization 
at some temperature $T_{s}(=\beta_{s}^{-1})$ 
as follows. 
\begin{eqnarray}
m_{0} & \equiv & \frac{1}{N}\sum_{i}\xi_{i} \nonumber \\
\mbox{} & = & \frac{{\rm tr}_{\xi}
\xi {\rm e}^{2m_{0}\beta_{s}\xi-\beta_{s}\xi^{2}}}
{{\cal Z}_{s}}.
\end{eqnarray}
We should be in mind that 
for the infinite range Q-Ising model, 
the properties of the 
macroscopic quantities of the system 
are completely determined by $m_{0}$. 
In FIG. \ref{fig1}, we plot the magnetization $m_{0}$ as a function 
of source temperature $T_{s}$ for $Q=3$ and $Q=4$. 
We see that for the $Q=3$ case the three states $m_{0}=1,2$ and 
$3$ are degenerated at $T_s=0$, while at finite temperature 
the middle state $m_{0}=2$ becomes globally stable  
and $m_{0}=1,3$ are degenerated locally stable. 
At high temperature regime $T_{s}\rightarrow \infty$, each spin takes 
all the values with same probability $1/3$ and thus the corresponding 
magnetization is $m_{0}=(1+2+3)/3=2$. 
The transition between the ferro-magnetic 
phase and the para-magnetic phase occurs at $T_{c}\,{\sim}\,1.0$.  
In the same way as the case of 
$Q=3$, for $Q=4$, the four states $m_{0}=1,2,3$ and 
$4$ are degenerated at $T_{s}=0$, and 
the middle two states $m_{0}=2$ and $3$ 
become globally stable for $T_{s}>0$ 
($m_{0}=1,4$ are degenerated locally stable states). 
The para-magnetic state is specified by the 
magnetization $m_{0}=(1+2+3+4)/4=2.5$ and 
the ferro-para  transition occurs at $T_{c}\,{\sim}\,1.78$. 
For this original image, in order to 
investigate the average performance of the 
MPM estimation, we should calculate $D_{H}$ in 
terms of statistical mechanics of the 
spin system $\{\sigma\}$ with quenched disorder 
$\{\tau\},\{J\}$ and $\{\xi\}$. 
For this purpose, we calculate the averaged 
free energy of the 
system described by ${\cal H}_{\rm eff}$ 
[Eq. (\ref{effectHam1})] with assistant of 
the replica method; 
\begin{eqnarray}
[{\log}{\cal Z}]_{\{\tau\},\{J\},\{\xi\}} & = & 
\displaystyle{\lim_{n\rightarrow 0}}
\frac{[{\cal Z}^{n}]_{\{\tau\},\{J\},\{\xi\}}-1}{n}
\end{eqnarray}
with 
\begin{eqnarray}
{\cal Z} & = & {\rm tr}_{\{\sigma\}}
{\exp}(-\beta {\cal H}_{\rm eff}),
\end{eqnarray}
which consist in replacing the 
quenched average of a single system 
with an annealed average of $n$ 
replicated systems (letting $n\rightarrow 0$ at the 
end).  
Assuming a replica symmetric ansatz   
and by using the saddle point method, 
the order parameters are
given by the following coupled equations;   
\begin{eqnarray}
m & = & [\langle \sigma_{i}^{\alpha} 
\rangle_{\beta_{d},h}]_{\{\tau\},\{J\},\{\xi\}}=  
{\rm tr}_{\xi}{\cal Q}(\xi)
\int_{-\infty}^{\infty} \!\!{\cal D}x\,
{\cal B}(x,\xi)
\label{m_rs} \\
t & = & [\xi_{i} \langle \sigma_{i}^{\alpha} 
\rangle_{\beta_{d},h}]_{\{\tau\},\{J\},\{\xi\}}= 
{\rm tr}_{\xi} \xi {\cal Q}(\xi)
\int_{-\infty}^{\infty} \!\!{\cal D}x\,{\cal B}(x,\xi)
\label{t_rs}\\
q & = & [\langle \sigma_{i}^{\alpha} 
\rangle_{\beta_{d},h}
\langle \sigma_{i}^{\beta} 
\rangle_{\beta_{d},h}]_{\{\tau\},\{J\},\{\tau\}}=
{\rm tr}_{\xi} {\cal Q}(\xi)
\int_{-\infty}^{\infty} \!\!{\cal D}x\,
[{\cal B}(x,\xi)]^{2}
\label{q_rs}\\
w & = & [\langle (\sigma_{i}^{\alpha})^{2} 
\rangle_{\beta_{d},h}]_{\{\tau\},\{J\},\{\xi\}}=
{\rm tr}_{\xi}
{\cal Q}(\xi)
\int_{-\infty}^{\infty} \!\!{\cal D}x\,
{\cal C}(x,\xi)
\label{w_rs}.  
\end{eqnarray}
where we should remember that 
the brackets 
$\langle \cdots \rangle_{\beta_{d},h}$ 
and $[\cdots]_{\{\tau\},\{J\},\{\xi\}}$ 
are defined by Eq. (\ref{f_ave1}) and 
Eq. (\ref{f_ave2}), respectively. 
In the above expressions, 
${\cal D}x\,\equiv\,(dx/\sqrt{2\pi}){\rm e}^{-x^{2}/2}$  
is the usual Gaussian measure and 
we defined  
\begin{eqnarray}
{\cal B}(x,\xi) & \equiv & 
\frac{{\rm tr}_{\sigma}
{\sigma}\,{\exp}\left[U\sigma-V\sigma^{2}
\right]}
{ {\rm tr}_{\sigma}
{\exp}\left[U\sigma-V\sigma^{2}
\right]} \\
{\cal C}(x,\xi) & \equiv & 
\frac{{\rm tr}_{\sigma}
{\sigma}^{2}\,{\exp}\left[
U\sigma-V\sigma^{2}
\right]}
{{\rm tr}_{\sigma}
{\exp}\left[
U\sigma-V\sigma^{2}
\right]}  \\
{\cal Q}(\xi) & \equiv & 
\frac{{\rm e}^{2 m_{0}\beta_{s}\xi-\beta_{s}\xi^{2}}}
{{\cal Z}(\beta_{s})}
\end{eqnarray}
with 
\begin{eqnarray}
U/2 &  \equiv  & 
(\tau_{0}h+\beta_{J}J_{0}t)\xi +m\beta_{d}
+x\sqrt{\tau^{2}h^{2}+\beta_{J}^{2}J^{2}q} \\  
V & \equiv &  h+\beta_{d}+\beta_{J}w.
\end{eqnarray}
Using the same way as the derivation 
of the order parameters, the averaged 
Hamming distance Eq. (\ref{aveHam1}) is 
calculated and reads 
\begin{eqnarray}
D_{\rm H}(\beta_{d},h,\beta_{J})  = 
{\rm tr}_{\xi}
{\cal Q}(\xi)
\int_{-\infty}^{\infty} \!\!{\cal D}x\,\left[
\xi-{\Omega}({\cal B}(x,\xi))
\right]^{2}. 
\end{eqnarray}
It is straightforward to check that the above equations coincide with those in 
\cite{NW} for $Q=2$ case.
To keep things easy, we first assume that there 
is no glassy term in our 
decoding process, {\it i.e.} $\beta_J=0$.
In FIG. 2, we plot the Hamming distance 
as function of the decoding temperature for $Q=3$ and $T_s=0.75$. The minimum 
is reached at $T_d=0.75$. The same for $Q=4$ in FIG. \ref{static_Q4}. 
It can be shown numerically, at least 
for $Q=3$ and $Q=4$, that, 
given the original image at temperature $T_s$, just below the transition 
temperature, the optimal decoding temperature $T_d^{\rm (opt)}=T_s$. 
The same relation turns out to be satisfied for  black or white IR \cite{NW} 
(which corresponds to $Q=2$), differently from 
GSIR by the Potts model \cite{CI}. 
In order to compare the 
performance of the MPM estimate with 
that of the MAP one, 
we first investigate the scaled field 
$H\,\equiv\,h/\beta_{d}$ 
dependence of the MAP estimate. 
The MAP estimate 
is obtained by controlling the 
temperature as $T_{d}\rightarrow 0$ with 
keeping the scaled field $H$ constant. 
Therefore, the Hamming distance 
for the MAP estimate should depends on $H$. 
In FIG. 4 {\sf (a)}, 
we plot the Hamming distance of the MAP 
estimate as a function of $H$. 
In this figure, we set $Q=3, T_{s}=0.75$ and 
$\tau=\tau_{0}=1.0$. 
We see that the Hamming distance 
takes its minimum at 
$H_{\rm opt}=\tau_{0}/2\tau^{2}\beta_{s}=0.375$, 
namely, 
$D_{\rm H}(T_{d}=0,H)\, \geq \,D_{\rm H}(T_{d}=0,
\tau_{0}/2\tau^{2}\beta_{s})$. 
This optimal value of the scaled 
field $H=h/\beta_{d}$ 
is obtained when we set 
$P(\{\tau\}|\{\xi\})=P(\{\tau\}|\{\sigma\})$ and 
$P_{s}(\{\xi\})=P_{d}(\{\sigma\})$, 
that is to say, 
$\beta_{d}=\beta_{s}$ and $h=\tau_{0}/2\tau^{2}$. 
In FIG. 4 {\sf (b)}, 
we increase the temperature $T_{d}$ 
with $H=H_{\rm opt}$ and plot the Hamming distance 
$D_{H}(T_{d},H_{\rm opt})$ as a 
function of $T_{d}$. 
This figure shows that 
$D_{\rm H}(T_{d}, H_{\rm opt})$ 
takes its maximum at $T_{d}=T_{s}=0.75$. 
Therefore, 
we conclude that the MPM estimate 
achieves the lowest Hamming distance 
which can not be obtained by the MAP estimation. 
In FIG. 4 {\sf (b)}, 
we plot the $D_{\rm H}(T_{d},H)$ 
for several values of $H$. 
From this figure, we see that as long as we choose $H$ so as to satisfy 
$H\, {\geq}\, H_{\rm opt}=\tau_{0}/2\tau^{2}\beta_{s}$, 
the minimum value of the Hamming distance does not change. 

In the limit of $T_{d}\rightarrow \infty$, 
each pixel takes $\sigma_{i}=1,2,3$ 
with same probability $1/3$, and 
the local magnetization leads to 
$\langle \sigma_{i} \rangle =(1+2+3)/3=2$ for 
all pixels $i$. 
As the result, the Hamming distance 
in the high temperature limit takes 
\begin{eqnarray}
D_{\rm H}(T_{d}\rightarrow \infty)=
\frac{\sum_{\xi=1}^{3}
(2-\xi)^{2}{\rm e}^{2m_{0}\beta_{s}\xi-\beta_{s}\xi^{2}}}
{2 \sum_{\xi=1}^{3}
{\rm e}^{2m_{0}\beta_{s}\xi-\beta_{s}\xi^{2}}}\,{\sim}\,0.1726.
\end{eqnarray}
This asymptotic behavior is checked in FIG. 4 {\sf (b)}. 
We now switch on the product 
interaction, that is, $\beta_{J}\neq 0$  setting $T_d$ and $H$ 
at their optimal values. As clearly shown in FIG. \ref{glass}, the 
performance of the restoration is dramatically improved. In this 
figure, the point $\beta_J=0$ 
corresponds to the minimum of the Hamming distance in FIG. \ref{static_Q3}.

\subsection{Dynamics \label{Dynamics}}
An important and interesting problem is to determine the basin 
of attraction of the Hamming distance $D_{\rm H}(t)$. 
In fact, because of the presence of locally stable states,
the final state of the decoding process is strongly dependent upon the initial 
condition of the dynamics.
In addition, as the number of local minima increases with the
the number of the gray-scale levels,  it becomes crucial to choose the initial 
state appropriately. 
However, as it is well known, it is difficult to treat 
the dynamics of spin system explicitly in finite 
dimension, especially, 
dynamics in two dimension 
which is the case of image restoration. 
In the previous section, we introduced the infinite range 
model and solved it analytically. 
Using this model, we derived the properties of 
image restoration and 
as we see in the next section, the results do not 
contradict qualitatively with 
the properties in two dimension. 
With this fact in mind, 
we also use the infinite range model 
to investigate the dynamical properties 
of image restoration. 
In the equilibrium limit 
$t \rightarrow \infty$, without glassy term, 
namely,  
$\beta_{J}=0$, 
the properties of image restoration in 
the infinite range model are 
completely written by magnetization $m$. 
Therefore, we assume that the dynamics 
of image restoration is also expressed by 
the time evolution of the 
magnetization $m(t)$. 
Therefore, we derive the differential equations with respect to 
the macroscopic variables, namely, 
$m(t)$ and $D_{\rm H}(t)$,  from the microscopic master equation. 
For the sake of simplicity, we restrict ourselves to the case 
without the glassy term. The master equation of our system leads to 
\begin{eqnarray}
\frac{d p_{t}(\{\sigma\})}{dt} = \sum_{k=1}^{N}
\sum_{\sigma_{k^{'}}=1}^{Q}
\left[
w({\sigma_{k^{'}}\rightarrow \sigma_{k}})p_{t}
(\{\sigma\}^{'})-
w({\sigma_{k}\rightarrow \sigma_{k^{'}}})
p_{t}(\{\sigma\})
\right]
\end{eqnarray}
with the following transition probability 
\begin{eqnarray}
w({\sigma_{k}\rightarrow \sigma_{k^{'}}}) =  
\frac{{\rm e}^{-(h+\beta_{d})\sigma_{k^{'}}^{2}
+2(m\beta_{d}+h\tau_{k})\sigma_{k^{'}}}}
{{\sum_{\sigma_{k^{'}}=1}^{Q}}
{\rm e}^{-(h+\beta_{d})\sigma_{k^{'}}^{2}
+2(m\beta_{d}+h\tau_{k})\sigma_{k^{'}}}
},
\end{eqnarray}
where  
we defined 
$\{\sigma\}  \equiv (\sigma_{1},{\cdots},\sigma_{k},
{\cdots},\sigma_{N})$ and 
$\{ \sigma \}^{'} \equiv  (\sigma_{1},{\cdots},\sigma_{k^{'}},
{\cdots},\sigma_{N})$.
By introducing the probability distribution of the macroscopic 
magnetization $m$  
,{\it viz.}
\begin{eqnarray}
{\cal P}_{t}(m) \equiv 
\sum_{\sigma}
p_{t}(\{ \sigma \})
{\delta}[m-m(\{ \sigma \})]
\end{eqnarray}
and after some algebra 
we obtain the following exact differential equation 
\begin{eqnarray}
\mbox{} & \mbox{} & \frac{d m}{dt} = -m +    
\sum_{\xi=1}^{Q}
{\cal Q}(\xi)\int_{-\infty}^{\infty}
 \!\!\!Dx \, 
{\cal B}(x,\xi){\biggr |}_{\beta_{J}=0}.
\label{dyn1}
\end{eqnarray}
The derivation of the above 
differential equation is 
reported in Appendix A.
The time evolution of the 
Hamming distance $D_{\rm H}(t)$ is obtained by 
substituting the time dependence of the 
magnetization $m(t)$ into $D_{\rm H}(m)$.   
In FIG. \ref{time_ev}, we plot the time evolutions of the $Q=3$ Hamming distance for 
$T_{d}=T_{s}$ {\sf (a)} and $T_{d}\neq T_{s}$ {\sf (b)}. 
From these figures, we see that 
if we choose the hyper-parameter $T_{d}$ so as to satisfy the  
relationship $T_{d}=T_{s}$, the Hamming distance converges to 
its optimal value for any initial condition. 
On the other hand, for $T_d\neq T_s$, 
there exists a threshold of the initial value of 
the Hamming distance $D_{\rm H}^{(c)}$  beyond which the 
flow $D_{\rm H}(t)$ does not converges to its optimal value. 
As  the dynamical equation (with respect to 
$m$) is exactly the same as the Time Dependent Ginsburg-Landau (TDGL) equation, 
that is, $dm/dt=-\partial f_{\rm RS}/\partial m$,  
the nature of the dynamics is intuitively  understood as 
a steepest descent to 
a local minimum of the free 
energy. 
In fact, from FIG. \ref{static_Q3} {\sf (a)},{\sf (b)} and {\sf (c)}, we
see that for $T_{d}\,<\, 0.35$, 
there exist local minima.Therefore if 
we fail to choose the initial condition appropriately, 
the Hamming distance converges to the non-optimal values. 
For practical situations,  the corrupted image corresponds to 
our initial state. For the case $T_{d}\,\neq\, T_{s}$, we calculate the 
Hamming distance $D_{\rm H}^{(1)}$ between the original image 
and the noised one, which reads
\begin{eqnarray}
D_{\rm H}^{(1)} & \equiv & 
\frac{1}{2N}\sum_{i}(\tau_{i}-\xi_{i})^{2} =  
\frac{\sum_{\xi=1}^{Q}[\tau^{2}+(\tau_{0}^{2}-1)\xi^{2}]
{\rm e}^{2 m_{0}\beta_{s}\xi-\beta_{s}\xi^{2}}}
{2{\cal Z}(\beta_{s})}.
\end{eqnarray}
In particular, for $\tau=\tau_{0}=1.0$, 
this leads to $D_{\rm H}^{(1)}=0.5$. 
From FIG.  5 {\sf (b)},  
we see that if we choose the corrupted image as an initial 
state, we destroy the observed corrupted image and 
the result is even worse.

The asymptotic expressions of the magnetization and 
the Hamming distance in the limit of $t\rightarrow \infty$ 
lead to  
\begin{eqnarray}
m & = & m_{*}+[m(t=0)-m_{*}]\,{\rm e}^{-\frac{t}{t_{0}}}, \\
D_{\rm H} & = & D_{\rm H}(m_{*})+\tilde{D}_{\rm H}\,
{\rm e}^{-\frac{t}{t_{0}}}
\end{eqnarray}
where $m_{*}$ is a solution of the saddle point equations 
Eqs. (\ref{m_rs})-(\ref{w_rs}) 
with $\beta_{J}=0$ 
in the previous section. 
The relaxation time $t_{0}$ is given as 
\begin{eqnarray}
\frac{1}{t_{0}}=
1+2\beta_{d}\int_{-\infty}^{\infty}
{\cal D}x{\cal Q}(\xi)\left[
\overline{\sigma^{2}}-(\overline{\sigma})^{2}
\right]
\end{eqnarray}
with 
\begin{eqnarray}
\overline{(\cdots)} \equiv 
\frac{\sum_{\sigma=1}^{Q}(\cdots)
{\rm e}^{2(m_{*}\beta_{d}+\tau h x+\tau_{0}h \xi)\sigma
-(h+\beta_{d})\sigma^{2}}}
{\sum_{\sigma=1}^{Q}
{\rm e}^{2(m_{*}\beta_{d}+\tau h x+\tau_{0}h \xi)\sigma
-(h+\beta_{d})\sigma^{2}}}, 
\end{eqnarray}
and $\tilde{D}_{\rm H}$ reads 
\begin{eqnarray}
\tilde{D}_{\rm H} & \equiv  &  
4\beta_{d}[m(t=0)-m_{*}]\int_{-\infty}^{\infty} {\cal D}x
[\xi-{\Omega}(\overline{\sigma})] \nonumber  \\
\mbox{} & \times & 
\sum_{k=1}^{Q} k 
\left[
{\delta}\left(
\overline{\sigma}-\frac{2k-1}{2}
\right)
-{\delta}\left(
\overline{\sigma}-\frac{2k+1}{2}
\right)
\right]
[\overline{\sigma^{2}}-(\overline{\sigma})^{2}]. 
\end{eqnarray}  
We plot the inverse relaxation time  
$1/t_{0}$ as a function of $T_{d}$ 
for the case of $Q=3,T_{s}=0.75, 
\tau_{0}=\tau=1.0$ in FIG. 7.
In this figure, we also plot the inverse 
relaxation time for several values of 
the scaled field $H$. 
We see that the inverse relaxation time $1/t_{0}$ takes 
its minimum at a finite temperature $T_{d}$.
However, the inverse relaxation time 
$1/t_{0}$ never reaches to zero, 
and the relaxation to the 
equilibrium state is exponential for all 
temperature $T_{d}$ regions. 
\section{Monte Carlo simulations \label{MC} }
So far, we worked under the assumption that all the pixels lay 
on an infinite dimensional grid, 
an approximation which enabled us to derive exact analytical formulas. 
In order to test the efficiency of the Q-Ising model on 
the more realistic case of 
two dimensional picture, in this section, 
we carry out Monte Carlo 
simulations at finite temperature 
on a real-world image with $Q=8$ gray-scale 
level [FIG. \ref{montecarlo_fig} {\sf (a)}] 
corrupted by a Gaussian noise with 
$\tau=1.2$ [FIG. \ref{montecarlo_fig} {\sf (b)}]. 
Here the interaction in effective 
Hamiltonian is now restricted to the nearest 
neighbors spins on two dimensional square lattice. 
As before,  we first study the Hamming distance 
without the glassy term. The resulting curves 
averaged over twenty Monte Carlo runs  are shown 
in FIG. (\ref{no_glass}) {\sf (a)} for three different 
values of the ratio 
$H=h/\beta_d$. The plots reflect indeed the mean 
field behavior of FIG. \ref{static_Q3} {\sf (d)} 
and \ref{static_Q4} {\sf (b)}. 
The corresponding restored picture at optimal values is 
shown in FIG. \ref{montecarlo_fig} 
{\sf (c)}. It is evident that the 
ferromagnetic term succeeds in eliminating 
the noised pixel, {\it i.e.} isolated ones, 
but at the same time it also smoothes out the small 
true details of the original picture. 
For this reason, by keeping fixed 
$T_d^{\rm (opt)}$ and $H_{\rm opt}$, 
we switch on the glassy term, namely, $\beta_{J} 
\neq 0$. Here we choose a slightly more general 
expression for $J_{ij}=(\xi_i-\xi_j)^2$ and therefore 
the extra term reads $\left([J_{ij}-(\sigma_i-\sigma_j)^2]^{2}/N\right)$.
The curves in FIG. \ref{no_glass} {\sf (b)} for twenty Monte Carlo 
runs, show an improvement of the recovered image. 
The restored image at the minimum $\beta_J^{\rm (opt)}$ 
is on the lower right corner of FIG. \ref{montecarlo_fig}. 
It is evident that  
the extra term preserves many of the small details 
of the original image, for example the white edge of the roof, 
which was blurred for the case of $\beta_J=0$.
\section{Iterative Algorithm (mean-field) }
\label{It}
The restoration by means of Monte Carlo methods 
is the result of a  statistical process 
which might take long time even for powerful 
computer as the size of the picture 
increases. Therefore we apply  
an iterative algorithm, proposed in \cite{Tanaka,Morita}, 
to our model.
Using the mean-field approximation with periodic 
boundary conditions for two dimensional 
square lattice of size $L_{1} \times L_{2}$, 
the recursion relations with respect to the local 
magnetization at a site $(i,j)$, namely, $m_{ij}$ lead to  
\begin{eqnarray}
m_{ij}^{(t+1)}=
\frac{{\rm tr}_{\sigma}
\sigma {\rm e}^{\omega_{ij}^{(t)}(\sigma)}}
{ {\rm tr}_{\sigma}
{\rm e}^{\omega_{ij}^{(t)}(\sigma)}},
\label{recur1}
\end{eqnarray}
\begin{eqnarray}
{\omega}_{ij}^{(t)}(\sigma)=
\left\{
J[m_{i,j+1}^{(t)}+m_{i,j-1}^{(t)}
+m_{i+1,j}^{(t)}+m_{i-1,j}^{(t)}]
+2h\tau_{ij}\right\}
{\sigma}-(2J+h)\sigma^{2}, 
\label{recur2}
\end{eqnarray}
\begin{eqnarray}
m_{i,j+L_{2}}=m_{ij},\,\,\,\,\,\,\,\,\,m_{i+L_{1},j}=m_{ij}.
\label{recur3}
\end{eqnarray}
where ${\rm tr}_{\sigma}(\cdots)$ 
means the sum with respect to 
the gray scale levels, namely, $\sigma=1,2,{\cdots},Q$. 
The details of the derivation by using 
a variational principle are  
reported in Appendix B.

Then, we obtain the estimate 
of the pixel $\xi_{ij}$, namely, 
$\Omega(m_{ij})$ by 
solving the above non-linear maps until appropriate  
error tolerance is satisfied. 
In order to investigate 
its performance,  
we introduce the following 
three measures 
\begin{eqnarray}
D_{\rm H} & \equiv & \frac{1}{2L_{1}L_{2}}
\sum_{i=1}^{L_{1}}
\sum_{j=1}^{L_{2}}
[\Omega(m_{ij})-
\xi_{ij}]^{2} \\
D_{\rm H}^{(1)} & \equiv & 
\frac{1}{2L_{1}L_{2}}
\sum_{i=1}^{L_{1}}
\sum_{j=1}^{L_{2}}
[\Omega(m_{ij})-
\tau_{ij}]^{2} \\
D_{\rm H}^{(2)} & \equiv & 
\frac{1}{2L_{1}L_{2}}
\sum_{i=1}^{L_{1}}
\sum_{j=1}^{L_{2}}
[\tau_{ij}-
\xi_{ij}]^{2}   
\end{eqnarray}
where 
$D_{\rm H}$ , $D_{\rm H}^{(1)}$ and 
$D_{\rm H}^{(2)}$ are  
distances between 
the original image $\xi_{ij}$ and 
the restored one $\Omega(m_{ij})$, 
the corrupted image $\tau_{ij}$ and 
the restored one, 
the corrupted image and the 
original one, respectively. 
We choose  
$Q=8, L_{1}=L_{2}=128$ and $h=1.0$ and solve the 
recursion relations  
(\ref{recur1}),(\ref{recur2}) and (\ref{recur3}) 
until the error 
\begin{eqnarray}
{\varepsilon}^{(t)}
\,\equiv\,\frac{1}{2L_{1}L_{2}}\sum_{i=1}^{L_{1}}\sum_{j=1}^{L_{2}}
|m_{ij}^{(t+1)}-m_{ij}^{(t)}|
\end{eqnarray}
becomes smaller than $10^{-5}$.
We list the results in FIG. 8 (``girl'') and 
FIG. 9 (``chair''). 
The original images are 
degraded by the 
Gaussian noise 
with a standard deviation $\tau=2.2$.
This standard deviation gives the Hamming 
distance between the original 
image and degraded one $D_{\rm H}^{(2)}\,{\sim}\,0.205$.
Obviously, the picture ``chair'' 
contains much more edges than 
the picture ``girl''. 
Therefore, one of the our aims 
of this demonstration is 
to check to what extent our 
model can detect the edge parts in the real-world picture. 
In FIG. 10, we plot the Hamming distance $D_{\rm H}$ 
[{\sf (a)}] and $D_{H}^{(1)}$ [{\sf (b)}].
We choose the degraded image as an 
initial set of the pixels and investigated 
the $J$-dependence of the 
Hamming distance. 
We see that 
the performance of 
the algorithm for the ``girl'' picture 
is much better than that of 
the ``chair'' picture. 
This is because the smoothness term in the 
effective Hamiltonian [Eq. (\ref{effHam0})] 
is quadratic and it is hard to detect 
the edges in the ``chair'' picture. 
For both pictures, 
the optimal performance is achieved
 around the parameter $T_{d}=1/J=1/1.8\,{\sim}\,0.56$.
This value is not so different from the parameter 
which was obtained in Monte Carlo simulations [see FIG. 6 {\sf (a)}]. 
Of course, from a practical point of 
view, it is possible to stop the 
Monte Carlo simulation and not to 
wait the convergence to 
the equilibrium state precisely. 
Then, we may regard the snap-shot as the 
restored image if the performance is not so bad. 
However, in the mean-field approximation we 
constructed here, the 
convergence of the iteration is guaranteed. 
\section{Summary and Discussions} 
In this paper we investigated the 
efficiency of the Q-Ising model 
for image restoration problem, when the original image 
is affected by Gaussian noise. 
By introducing the 
infinite range model, we gave an analytical expression for the 
Hamming distance, which is shown to reach its minimum
at some finite temperature. 
We found that 
the optimal temperature for 
the GSIR using the Q-Ising model 
coincides with 
the source 
temperature in 
contrast to the chiral Potts case \cite{CI}.
We also found that as in the Ising and Potts spin cases, 
the presence of a parity-check-{\it like} term 
greatly increasing the performance of the GSIR process. 
Although for practical restorations of images, one wouldn't like 
to smooth out two points far away, the mean 
field results provide a remarkable eye-guide
for a short-range version of the effective Hamiltonian as 
confirmed by Monte Carlo simulation on two dimensional pictures. 
From a dynamical point of view, we also obtained the time 
evolution of the Hamming distance analytically and 
found the critical initial Hamming 
distance beyond which the flow 
does not converges to its optimal value.
We show the dynamical equation we obtained is 
exactly same as the TDGL equation. 
Therefore, the destination of 
the dynamics is one of the local minima of the free 
energy, and if we fail to 
select the initial condition, 
the dynamics converges to 
the local minimum which 
does not give 
the minimum of the Hamming 
distance. 

Recently, 
Skantzos and Coolen \cite{SC}, 
reported the synchronous 
dynamics of 1-D and infinite range version of 
the random field Ising model. 
They found that the dynamics has 
much more rich behavior 
than 
the sequential (Glauber) dynamics. 
Therefore, for our present model system, 
there is a possibility 
that if we consider 
the synchronous dynamics instead of 
the sequential one, the behavior of 
the dynamics may be different from 
the results we obtained here. 

Using the mean-field approximation, 
we also constructed the iterative 
algorithm which converges faster than 
the Monte Carlo simulation. 
We derived it 
from a variational principle of the 
free energy and demonstrate it for two types of 
the real-world pictures. 
From those results, 
we concluded that 
we need some extra term 
which detect the edges if 
the picture has a lot of edges. 
We suppose that the glassy term 
we introduced in the infinite range model 
may play this role. 
This will be achieved by means of 
the TAP like mean-field approximation. 

The authors acknowledge  Profs. 
Hidetoshi Nishimori, Kazuyuki Tanaka, and 
Desire Bolle'  for fruitful discussions and useful comments. 
We also thank Dr. Masato Okada for 
useful discussions about the dynamics of 
disordered systems.

One of the authors (J. I.) was partially   
supported by the Ministry of Education, 
Science, Sports and Culture, 
Grant-in-Aid for 
Encouragement of Young Scientists, 
No. 11740225, 1999-2000.  
\appendix
\section{Derivation of the flow of magnetization}

In this Appendix, we derive 
the differential equation with respect to 
macroscopic order parameter $m$ from microscopic master equation 
for the infinite range version of the Q-Ising model. 
For simplicity, 
we consider the case of no-parity check term 
$\beta_{J}=0$.
For the Q-Ising model, the effective 
Hamiltonian is given as 
\begin{eqnarray}
{\cal H}_{\rm eff}  =  
\frac{\beta_{d}}{2N}
\sum_{ij}(\sigma_{i}-\sigma_{j})^{2}
+h\sum_{i}(\sigma_{i}-\tau_{i})^{2} \equiv {\cal H}(\sigma_{k}).  
\end{eqnarray}
Therefore, the energy difference 
due to the local spin change 
$\sigma_{k}\,\rightarrow\,\sigma_{k^{'}}$ , namely, 
$\Delta E \equiv {\cal H}(\sigma_{k^{'}})-{\cal H}(\sigma_{k})$ 
is calculated in terms of 
the above Hamiltonians 
${\cal H}(\sigma_{k})$ and ${\cal H}(\sigma_{k^{'}})$ as 
follows. 
\begin{eqnarray}
{\Delta}E  =  
(h+\beta_{d})(\sigma_{k^{'}}^{2}-\sigma_{k}^{2})-
2\beta_{d}m(\sigma_{k^{'}}-\sigma_{k})-2h
(\sigma_{k^{'}}-\sigma_{k})\tau_{k},
\end{eqnarray}
where we used the expression of the magnetization 
\begin{eqnarray}
m=\frac{1}{N}\sum_{j}\sigma_{j}.
\end{eqnarray}
Then, the transition probability $w(\sigma_{k}\rightarrow \sigma_{k^{'}})$ 
is given by 
\begin{eqnarray}
w(\sigma_{k}\rightarrow \sigma_{k^{'}}) & = & 
\frac{{\rm e}^{-\Delta E}}
{{\sum_{\sigma_{k^{'}}=1}^{Q}}{\rm e}^{-\Delta E}} \nonumber \\
\mbox{} & = & 
\frac{{\rm e}^{-(h+\beta_{d})\sigma_{k^{'}}^{2}
+2(m\beta_{d}+h\tau_{k})\sigma_{k^{'}}}}
{{\sum_{\sigma_{k^{'}}=1}^{Q}}
{\rm e}^{-(h+\beta_{d})\sigma_{k^{'}}^{2}
+2(m\beta_{d}+h\tau_{k})\sigma_{k^{'}}}
}.
\end{eqnarray}
For this transition probability, 
the master equation leads to 
\begin{eqnarray}
\frac{d}{dt}\,p_{t}(\{\sigma \}) & = & 
\sum_{k=1}^{N}\sum_{\sigma_{k^{'}}=1}^{Q}
\left[
w(\sigma_{k^{'}}\rightarrow \sigma_{k})p_{t}
(\{ \sigma \}^{'})-
w(\sigma_{k}\rightarrow \sigma_{k^{'}})
p_{t}(\{ \sigma \})
\right].
\end{eqnarray}
Here we introduce the 
following macroscopic probability 
\begin{eqnarray}
{\cal P}_{t}(m) \equiv 
\sum_{\{\sigma\}}
p_{t}(\{ \sigma \})
{\delta}[m-m(\{ \sigma \})], 
\end{eqnarray}
and consider the derivative 
of ${\cal P}_{t}(m)$ with respect to $t$, that is,  
\begin{eqnarray}
\frac{d}{dt}
{\cal P}_{t}(m) & = & 
\sum_{\{\sigma \}}
\frac{d p_{t}(\{ \sigma \})}{dt}
{\delta}[m-m(\{ \sigma \})] \nonumber \\
\mbox{} & = & 
\sum_{\{ \sigma \}}
\left\{
\sum_{k=1}^{N}
\sum_{\sigma_{k^{'}}=1}^{Q}
\left[
w(\sigma_{k^{'}}{\rightarrow}\sigma_{k})
p_{t}(\{ \sigma \})-
w(\sigma_{k}{\rightarrow}\sigma_{k^{'}})
p_{t}(\{ \sigma \})
\right]
\right\}
{\delta}[m-m(\{ \sigma \})] \nonumber \\
\mbox{} & = & 
\sum_{\{ \sigma \}}
\sum_{k=1}^{N}
\sum_{\sigma_{k^{'}}=1}^{Q}
w(\sigma_{k}{\rightarrow}\sigma_{k^{'}})
p_{t}(\{ \sigma \})
\left\{
{\delta}\left[m-m(\{ \sigma \})+
\frac{1}{N}(\sigma_{k}-\sigma_{k^{'}})
\right]-
{\delta}[
m-m(\{ \sigma \})]
\right\} \nonumber \\
\mbox{} & = & 
\frac{\partial}{\partial m}
\left\{
\sum_{\{ \sigma \}}
\sum_{k=1}^{N}
\sum_{\sigma_{k^{'}}=1}^{Q}
w(\sigma_{k}{\rightarrow}\sigma_{k^{'}})
p_{t}(\{ \sigma \})
{\delta}[
m-m(\{ \sigma \})]
\frac{1}{N}(\sigma_{k}-\sigma_{k^{'}})
\right\} \nonumber \\
\mbox{} & = & 
\frac{\partial}{\partial m}
{\Biggr \{}
\sum_{\{ \sigma \}}
p_{t}(\{ \sigma \})
\sum_{k=1}^{N}
\sum_{\sigma_{k^{'}}=1}^{Q}
\left[
\frac{{\rm e}^{-(h+\beta_{d})\sigma_{k^{'}}^{2}
+2(m\beta_{d}+\tau_{k} h)\sigma_{k^{'}}}}
{{\sum_{\sigma_{k^{'}=1}}^{Q}}
{\rm e}^{-(h+\beta_{d})\sigma_{k^{'}}^{2}
+2(m\beta_{d}+\tau_{k} h)\sigma_{k^{'}}}}
\right] \nonumber \\
\mbox{} & \times & 
\frac{1}{N}(\sigma_{k}-\sigma_{k^{'}})
{\delta}(\sigma_{k}-\sigma_{k^{'}})
{\Biggr \}} \nonumber \\
\mbox{} & = & 
\frac{\partial}{\partial m}
\sum_{\{ \sigma \}}
p_{t}(\{ \sigma \})
\sum_{k=1}^{N}
{\Biggr \{}
\sum_{\sigma_{k^{'}}=1}^{Q}
\frac{\sigma_{k}}{N}
\left[
\frac{{\rm e}^{-(h+\beta_{d})\sigma_{k^{'}}^{2}
+2(m\beta_{d}+\tau_{k} h)\sigma_{k^{'}}}}
{{\sum_{\sigma_{k^{'}=1}}^{Q}}
{\rm e}^{-(h+\beta_{d})\sigma_{k^{'}}^{2}
+2(m\beta_{d}+\tau_{k} h)\sigma_{k^{'}}}}
\right] \nonumber \\
\mbox{} & - & 
\sum_{\sigma_{k^{'}}=1}^{Q}
\frac{\sigma_{k^{'}}}{N}
\left[
\frac{{\rm e}^{-(h+\beta_{d})\sigma_{k^{'}}^{2}
+2(m\beta_{d}+\tau_{k} h)\sigma_{k^{'}}}}
{{\sum_{\sigma_{k^{'}=1}}^{Q}}
{\rm e}^{-(h+\beta_{d})\sigma_{k^{'}}^{2}
+2(m\beta_{d}+\tau_{k} h)\sigma_{k^{'}}}}
\right]
{\Biggr \}}{\delta}[m-m(\{ \sigma \})] \nonumber \\
\mbox{} & = & 
\frac{\partial}{\partial m}
\sum_{\{ \sigma \}}
p_{t}(\{ \sigma \})
{\Biggr \{}
\frac{1}{N}
\left[\sum_{k=1}^{N}\sigma_{k}
-\sum_{k=1}^{N}
\frac{{\sum_{\sigma_{k^{'}}=1}^{Q}}
\sigma_{k^{'}}{\rm e}^{-(h+\beta_{d})\sigma_{k^{'}}^{2}
+2(m\beta_{d}+\tau_{k} h)\sigma_{k^{'}}}}
{{\sum_{\sigma_{k^{'}=1}}^{Q}}
{\rm e}^{-(h+\beta_{d})\sigma_{k^{'}}^{2}
+2(m\beta_{d}+\tau_{k} h)\sigma_{k^{'}}}}
\right]
{\Biggr \}} \nonumber \\
\mbox{} & \times & {\delta}[m-m(\{ \sigma \})].
\label{dpdt}
\end{eqnarray}
Here we should notice that 
\begin{eqnarray}
\mbox{} & \mbox{} & \frac{1}{N}
\sum_{k=1}^{N}
\frac{{\sum_{\sigma_{k^{'}}=1}^{Q}}
\sigma_{k^{'}}{\rm e}^{-(h+\beta_{d})\sigma_{k^{'}}^{2}
+2(m\beta_{d}+\tau h)\sigma_{k^{'}}}}
{{\sum_{\sigma_{k^{'}=1}}^{Q}}
{\rm e}^{-(h+\beta_{d})\sigma_{k^{'}}^{2}
+2(m\beta_{d}+\tau h)\sigma_{k^{'}}}} \nonumber \\
\mbox{} & = & 
\frac{{\sum_{\xi=1}^{Q}}
{\rm e}^{2 m_{0}\beta_{s}\xi-\beta_{s}\xi^{2}}}
{{\cal Z}(\beta_{s})}
\int_{-\infty}^{\infty} {\cal D}x 
\left[
\frac{\sum_{\sigma=1}^{Q}\sigma
{\rm e}^{-(h+\beta_{d})\sigma^{2}
+2(m\beta_{d}+h\tau x+h\tau_{0}\xi)\sigma}}
{\sum_{\sigma=1}^{Q}
{\rm e}^{-(h+\beta_{d})\sigma^{2}
+2(m\beta_{d}+h\tau x+h\tau_{0}\xi)\sigma}}
\right]
\end{eqnarray}
should be hold due to the self-averaging properties 
in the thermo-dynamical limit $N\rightarrow \infty$.
Substituting this expression  
into Eq. (\ref{dpdt}), we obtain  
\begin{eqnarray}
\frac{d}{dt}
{\cal P}_{t}(m) & = & 
\frac{\partial}{\partial m}m
\sum_{\{ \sigma \}}
p_{t}(\{ \sigma \}){\delta}
[m-m(\{ \sigma \})] 
-\frac{\partial}{\partial m}
\sum_{\{ \sigma \}}
p_{t}(\{ \sigma \})
{\delta}[m-m(\{ \sigma \})] \nonumber \\
\mbox{} & \times & 
\frac{{\sum_{\xi=1}^{Q}}
{\rm e}^{2 m_{0}\beta_{s}\xi-\beta_{s}\xi^{2}}}
{{\cal Z}(\beta_{s})}
\int_{-\infty}^{\infty} {\cal D}x 
\left[
\frac{\sum_{\sigma=1}^{Q}\sigma
{\rm e}^{-(h+\beta_{d})\sigma^{2}
+2(m\beta_{d}+h\tau x+h\tau_{0}\xi)\sigma}}
{\sum_{\sigma=1}^{q}
{\rm e}^{-(h+\beta_{d})\sigma^{2}
+2(m\beta_{d}+h\tau x+h\tau_{0}\xi)\sigma}} 
\right] \nonumber \\
\mbox{} & = & 
\frac{\partial}{\partial m}
m{\cal P}_{t}(m) \nonumber \\
\mbox{} & - & 
\frac{\partial}{\partial m}
{\cal P}_{t}(m)
\frac{{\sum_{\xi=1}^{Q}}
{\rm e}^{2 m_{0}\beta_{s}\xi-\beta_{s}\xi^{2}}}
{{\cal Z}(\beta_{s})}
\int_{-\infty}^{\infty} {\cal D}x 
\left[
\frac{{\sum_{\sigma=1}^{Q}}\sigma
{\rm e}^{-(h+\beta_{d})\sigma^{2}
+2(m\beta_{d}+h\tau x+h\tau_{0}\xi)\sigma}}
{{\sum_{\sigma=1}^{Q}}
{\rm e}^{-(h+\beta_{d})\sigma^{2}
+2(m\beta_{d}+h\tau x+h\tau_{0}\xi)\sigma}} 
\right] \nonumber \\
\mbox{} & = & 
\frac{\partial}{\partial m}
{\biggr \{}
{\cal P}_{t}(m)
{\biggr (}
m-\frac{{\sum_{\xi=1}^{Q}}
{\rm e}^{2 m_{0}\beta_{s}\xi-\beta_{s}\xi^{2}}}
{{\cal Z}(\beta_{s})} \nonumber \\
\mbox{} & \times & 
\int_{-\infty}^{\infty} {\cal D}x 
\left[
\frac{\sum_{\sigma=1}^{Q}\sigma
{\rm e}^{-(h+\beta_{d})\sigma^{2}
+2(m\beta_{d}+h\tau x+h\tau_{0}\xi)\sigma}}
{\sum_{\sigma=1}^{Q}
{\rm e}^{-(h+\beta_{d})\sigma^{2}
+2(m\beta_{d}+h\tau x+h\tau_{0}\xi)\sigma}}
\right]
{\biggr )}{\biggr \}}.
\label{dpdt2}
\end{eqnarray}
Multiplying 
$m$ and 
substituting ${\cal P}_{t}(m)={\delta}
[m-m(t)]$ to 
the left hand side of the above Eq. (\ref{dpdt2}), we obtain  
\begin{eqnarray}
\int_{-\infty}^{\infty} mdm\frac{d}{dt}{\delta}[m-m(t)]=
\frac{d}{dt}\int_{-\infty}^{\infty} m\,dm\,{\delta}[m-m(t)]=\frac{d m}{dt}.
\label{dpdt2left}
\end{eqnarray}
Using the same way as the left hand side of 
the Eq. (\ref{dpdt2}), the right-hand side of 
Eq. (\ref{dpdt2}) leads to  
\begin{eqnarray}
\mbox{} & \mbox{} & \int_{-\infty}^{\infty} mdm \frac{\partial}{\partial m}
{\biggr [}
\delta[m-m(t)]
{\biggr (}
m - 
\frac{{\sum_{\xi=1}^{Q}}
{\rm e}^{2 m_{0}\beta_{s}\xi-\beta_{s}\xi^{2}}}
{{\cal Z}(\beta_{s})} \nonumber \\
\mbox{} & \times & \int_{-\infty}^{\infty} {\cal D}x 
\left[
\frac{{\sum_{\sigma=1}^{Q}}\sigma
{\rm e}^{-(h+\beta_{d})\sigma^{2}
+2(m\beta_{d}+h\tau x+h\tau_{0}\xi)\sigma}}
{{\sum_{\sigma=1}^{Q}}
{\rm e}^{-(h+\beta_{d})\sigma^{2}
+2(m\beta_{d}+h\tau x+h\tau_{0}\xi)\sigma}}
\right]
{\biggr )}
{\biggr ]} \nonumber \\
\mbox{} & = & 
-\int_{-\infty}^{\infty} dm \,\delta[m-m(t)]
{\biggr \{}
m-
\frac{{\sum_{\xi=1}^{Q}}
{\rm e}^{2 m_{0}\beta_{s}\xi-\beta_{s}\xi^{2}}}
{{\cal Z}(\beta_{s})} \nonumber \\ 
\mbox{} & \times & \int_{-\infty}^{\infty} {\cal D}x
\left[ 
\frac{{\sum_{\sigma=1}^{Q}}\sigma
{\rm e}^{-(h+\beta_{d})\sigma^{2}
+2(m\beta_{d}+h\tau x+h\tau_{0}\xi)\sigma}}
{{\sum_{\sigma=1}^{Q}}
{\rm e}^{-(h+\beta_{d})\sigma^{2}
+2(m\beta_{d}+h\tau x+h\tau_{0}\xi)\sigma}}
\right]
{\biggr \}} \nonumber \\
\mbox{} & = & -m + 
\frac{{\sum_{\xi=1}^{Q}}
{\rm e}^{2 m_{0}\beta_{s}\xi-\beta_{s}\xi^{2}}}
{{\cal Z}(\beta_{s})}
\int_{-\infty}^{\infty} {\cal D}x
\left[ 
\frac{{\sum_{\sigma=1}^{Q}}\sigma
{\rm e}^{-(h+\beta_{d})\sigma^{2}
+2(m\beta_{d}+h\tau x+h\tau_{0}\xi)\sigma}}
{{\sum_{\sigma=1}^{Q}}
{\rm e}^{-(h+\beta_{d})\sigma^{2}
+2(m\beta_{d}+h\tau x+h\tau_{0}\xi)\sigma}}
\right]. 
\label{dpdt2right}
\end{eqnarray}
From Eqs. (\ref{dpdt2left}) and (\ref{dpdt2right}), 
we obtain the final form of 
the dynamical equation with respect to 
magnetization $m$ as 
\begin{eqnarray}
\frac{d m}{dt}= -m + 
\frac{{\sum_{\xi=1}^{Q}}
{\rm e}^{2 m_{0}\beta_{s}\xi-\beta_{s}\xi^{2}}}
{{\cal Z}(\beta_{s})}
\int_{-\infty}^{\infty} {\cal D}x 
\left[
\frac{{\sum_{\sigma=1}^{Q}}\sigma
{\rm e}^{-(h+\beta_{d})\sigma^{2}
+2(m\beta_{d}+h\tau x+h\tau_{0}\xi)\sigma}}
{{\sum_{\sigma=1}^{Q}}
{\rm e}^{-(h+\beta_{d})\sigma^{2}
+2(m\beta_{d}+h\tau x+h\tau_{0}\xi)\sigma}}
\right]. 
\label{app_dmdt}
\end{eqnarray}
We easily see that 
the above equation is exactly same as the 
Time Dependent Ginsburg-Landau (TDGL) equation 
which is derived from a steepest descent 
of the replica symmetric free energy, that is, 
$-{\partial f_{\rm RS}}/{\partial m}=dm/dt$.
We should also notice that 
in the limit of $t \rightarrow\infty$ and $dm/dt=0$, 
Eq. (\ref{app_dmdt})   
corresponds to the saddle point equation 
with respect to $m$ which was calculated by 
equilibrium statistical mechanics 
in Sec. IIA. 
\section{Variational Principle for the Q-Ising Model}
In Sec. VI,  
we introduced the recursion 
relations which determine the 
estimate of the original image 
in terms of mean-field approximation \cite{Tanaka,Morita}.
In this appendix,  
we show that these  
recursion relations 
Eqs. (\ref{recur1}), (\ref{recur2}) and 
(\ref{recur3}) can be derived from a  
variational principle. 

We consider the following optimization 
problem; 
\begin{eqnarray}
{\rm min}_{\rho}
\left\{
{\cal E}(\rho)-T{\cal S}(\rho)
\right\} 
\end{eqnarray}
\begin{eqnarray}
{\cal E}(\rho) & \equiv & \sum_{\{ \sigma \}}
{\cal H}(\{\sigma \})\,{\rho}(\{ \sigma \}) \\
{\cal S}(\rho) & \equiv & -\sum_{\{\sigma\}}
\rho(\{ \sigma \})\,
{\ln}\rho(\{ \sigma \}), 
\end{eqnarray}
where  Hamiltonian ${\cal H}$ is defined on the 
two dimensional square lattice of size 
$L_{1} \times L_{2}$ ($L_{1}=L_{2}=N$) as 
\begin{eqnarray}
{\cal H}(\{ \sigma \}) & = &  
\frac{J}{2}\sum_{ij}\{
(\sigma_{ij}-\sigma_{i,j+1})^{2}
+(\sigma_{ij}-\sigma_{i,j+1})^{2}
+(\sigma_{ij}-\sigma_{i+1,j})^{2}
+(\sigma_{ij}-\sigma_{i-1,j})^{2}
\} \nonumber \\
\mbox{} & + & h\sum_{ij}(\tau_{ij}-\sigma_{ij})^2, 
\end{eqnarray}
and  
${\cal E}$ and ${\cal S}$ 
correspond to 
the energy and entropy of the system, 
respectively. Then, 
we use the mean-field approximation, that is, 
\begin{eqnarray}
{\rho}(\{ \sigma \})\,\simeq\,
\prod_{ij}{\rho}_{ij}(\sigma_{ij}).
\end{eqnarray}
We should notice that 
for each pixel $(i,j)$, 
the following normalization 
condition should hold 
\begin{eqnarray}
\sum_{\sigma_{ij}=1}^{Q}
{\rho}_{ij}(\sigma_{ij})=1.
\label{normal}
\end{eqnarray}
Using the 
Lagrange multiplier $\lambda_{ij}$, 
we take into account the above 
normalization condition with 
respect to the marginal distribution,  and  
maximize the following 
functional  
\begin{eqnarray}
{\cal F} \equiv  {\cal E}
(\{ \sigma \})-T{\cal S}(\{ \sigma \})
+\sum_{ij}\lambda_{ij}\left(
\sum_{\sigma_{ij}=1}^{Q}
{\rho}_{ij}(\sigma_{ij})-1
\right).
\label{lagrangian}
\end{eqnarray}
The energy ${\cal E}$ and 
the entropy ${\cal S}$ of the 
system can be 
written explicitly as 
\begin{eqnarray}
{\cal E}  = 
\sum_{\sigma_{12}}
{\cdots}\sum_{\sigma_{ij}}
{\cdots}\sum_{\sigma_{kl}}
{\cdots}\sum_{\sigma_{N-1N}}
{\cal H}(\{ \sigma \})  
{\rho}_{12}(\sigma_{12})
{\cdots}{\rho}_{ij}(\sigma_{ij})
{\cdots}{\rho}_{kl}(\sigma_{kl})
{\cdots}{\rho}_{N-1N}(\sigma_{N-1N}),
\end{eqnarray}
\begin{eqnarray}
{\cal S} =   
 -\sum_{\sigma_{12}}\sum_{\sigma_{13}}
{\cdots}\sum_{\sigma_{ij}}
{\cdots}\sum_{\sigma_{kl}}
{\cdots}\sum_{\sigma_{N-1N}}
\prod_{ij}{\rho}_{ij}(\sigma_{ij})\sum_{ij}\,
{\ln}\,{\rho}_{ij}\,(\sigma_{ij}).
\end{eqnarray}
The derivative 
of the third term of 
Eq. (\ref{lagrangian}) 
with respect to $\rho_{ij}(\sigma_{ij})$ 
leads to $\sum_{\sigma_{ij}=1}^{Q}\lambda_{ij}$,  
therefore, we have  
$(\partial {\cal F}/\partial \rho_{ij})=0$ as 
\begin{eqnarray}
\frac{\partial {\cal F}}{\partial \rho_{ij}(\sigma_{ij})} & = &   
\sum_{\sigma_{ij}=1}^{Q}
{\biggr \{}
\sum_{\sigma_{12}}
{\cdots}\sum_{\sigma_{ij}}
{\cdots}\sum_{\sigma_{kl}}
{\cdots}\sum_{\sigma_{N-1N}}
{\cal H}(\{ \sigma \}){\rho}_{12}(\sigma_{12})
{\cdots}{\rho}_{kl}(\sigma_{kl})
{\cdots}{\rho}_{N-1N}(\sigma_{N-1N}) \nonumber \\
\mbox{} & + & T\,{\ln}\,{\rho}_{ij}
(\sigma_{ij})+T
+T\sum_{kl\neq ij}\sum_{\sigma_{kl}=1}^{Q}
\rho_{kl}(\sigma_{kl})
{\ln}\,\rho_{kl}(\sigma_{kl})
+\lambda_{ij}
{\biggr \}}=0.
\end{eqnarray}
This leads to 
\begin{eqnarray}
{\rho}_{ij}(\sigma_{ij})={\cal A}\,{\rm e}^{-\frac{1}{T}
\sum_{\sigma_{12}}
{\cdots}\sum_{\sigma_{ij}}
{\cdots}\sum_{\sigma_{kl}}
{\cdots}\sum_{\sigma_{N-1N}}
H(\{\sigma\})
{\rho}_{12}(\sigma_{12})
{\cdots}{\rho}_{kl}(\sigma_{kl})
{\cdots}{\rho}_{N-1N}(\sigma_{N-1N})}.
\label{marginal2}
\end{eqnarray}
Using the normalization condition (\ref{normal}), 
we obtain the factor 
\begin{eqnarray}
{\cal A} \equiv {\exp}\left(
-\frac{\lambda_{ij}}{T}-1-
\sum_{kl\neq ij}\sum_{\sigma_{kl}=1}^{Q}
\rho_{kl}(\sigma_{kl})
{\ln}\,\rho_{kl}(\sigma_{kl})
\right)
\end{eqnarray}
as 
\begin{eqnarray}
{\cal A}=
\left[\sum_{\sigma_{ij}=1}^{Q}
{\rm e}^{-\frac{1}{T}
\sum_{\sigma_{12}}
{\cdots}\sum_{\sigma_{ij}}
{\cdots}\sum_{\sigma_{kl}}
{\cdots}\sum_{\sigma_{N-1N}}
{\cal H}(\{ \sigma \})
{\rho}_{12}(\sigma_{12})
{\cdots}{\rho}_{kl}(\sigma_{kl})
{\cdots}{\rho}_{N-1N}(\sigma_{N-1N})}
\right]^{-1}.
\label{normal2}
\end{eqnarray}
From Eqs. (\ref{marginal2}) and (\ref{normal2}), 
the marginal distribution  $\rho_{ij}(\sigma_{ij})$ reads 
\begin{eqnarray}
{\rho}_{ij}(\sigma_{ij})=
\frac{
{\rm e}^{-\frac{1}{T}
\sum_{\sigma_{12}}
{\cdots}\sum_{\sigma_{ij}}
{\cdots}\sum_{\sigma_{kl}}
{\cdots}\sum_{\sigma_{N-1N}}
{\cal H}(\{ \sigma \})
{\rho}_{12}(\sigma_{12})
{\cdots}{\rho}_{kl}(\sigma_{kl})
{\cdots}{\rho}_{N-1N}(\sigma_{N-1N})}
}
{\sum_{\sigma_{ij}=1}^{Q}
{\rm e}^{-\frac{1}{T}
\sum_{\sigma_{12}}
{\cdots}\sum_{\sigma_{ij}}
{\cdots}\sum_{\sigma_{kl}}
{\cdots}\sum_{\sigma_{N-1N}}
{\cal H}(\{ \sigma \})
{\rho}_{12}(\sigma_{12})
{\cdots}{\rho}_{kl}(\sigma_{kl})
{\cdots}{\rho}_{N-1N}(\sigma_{N-1N})}
}.
\end{eqnarray}
In order to calculate 
the sum $\sum_{\sigma_{12}}{\cdots}
\sum_{\sigma_{NN-1}}(\cdots)$, 
we rewrite the Hamiltonian ${\cal H}(\{\sigma\})$ as 
\begin{eqnarray}
{\cal H}(\{\sigma \}) & = & 
-J(\sigma_{i,j+1}+\sigma_{i,j-1}
+\sigma_{i+1,j}+\sigma_{i-1,j})\sigma_{ij}
-2h\tau_{ij}\sigma_{ij}+(2J+h)(\sigma_{ij})^{2} \nonumber \\
\mbox{} & - & 
J\sum_{kl \neq ij}(\sigma_{k,l+1}+\sigma_{k,l-1}
+\sigma_{k+1,l}+\sigma_{k-1,l}){\sigma}_{kl}+
(2J+h)\sum_{kl \neq ij}(\sigma_{kl})^{2} \nonumber \\
\mbox{} & + & h\sum_{ij}(\tau_{ij})^{2}
-2h\sum_{kl \neq ij}
\tau_{kl}{\sigma}_{kl},  
\end{eqnarray}
and using the relations 
between the local magnetization and 
the marginal distribution, namely,  
$m_{i,j+1}= \sum_{\sigma_{i,j+1}=1}^{Q}
{\sigma}_{i,j+1}\,{\rho}(\sigma_{i,j+1})$, {\it etc.},
we obtain  
\begin{eqnarray}
\mbox{} & \mbox{} & 
-\frac{1}{T}
\sum_{\sigma_{12}}
{\cdots}\sum_{\sigma_{ij}}
{\cdots}\sum_{\sigma_{kl}}
{\cdots}\sum_{\sigma_{N-1N}}
{\cal H}(\{ \sigma \})
{\rho}_{12}(\sigma_{12})
{\cdots}{\rho}_{kl}(\sigma_{kl})
{\cdots}{\rho}_{N-1N}(\sigma_{N-1N}) \nonumber \\
\mbox{} & = & 
\frac{J}{T}(
m_{i,j+1}+m_{i,j-1}+m_{i+1,j}+m_{i-1,j})\sigma_{ij}+
\frac{2h}{T}\tau_{ij}\sigma_{ij}
-\frac{(2J+h)}{T}(\sigma_{ij})^{2} \nonumber \\
\mbox{} & - &    
\frac{1}{T}
\sum_{\{ \sigma \}{\in}I^{'}}
{\cal H}(\{ \sigma \})
{\rho}_{12}(\sigma_{12})
{\cdots}{\rho}_{kl}(\sigma_{kl})
{\cdots}{\rho}_{N-1N}(\sigma_{N-1N}). 
\label{factor}
\end{eqnarray}
where $I^{'}$ 
stands for a set of the sites except for 
$(i,j)$. 
\\
Using (\ref{factor}), we rewrite $\rho_{ij}(\sigma_{ij})$ as 
\begin{eqnarray}
{\rho}_{ij}(\sigma_{ij})=
\frac{{\rm e}^{\frac{1}{T}[
J(m_{i,j+1}+m_{i,j-1}+m_{i+1,j}+m_{i-1,j})
+2h\tau_{ij}]\sigma_{ij}-\frac{(2J+h)}{T}
(\sigma_{ij})^{2}}}
{\sum_{\sigma_{ij}=1}^{Q}
{\rm e}^{\frac{1}{T}[
J(m_{i,j+1}+m_{i,j-1}+m_{i+1,j}+m_{i-1,j})
+2h\tau_{ij}]\sigma_{ij}-\frac{(2J+h)}{T}
(\sigma_{ij})^{2}}}, 
\end{eqnarray}
where the factors 
\[
{\rm e}^{-\frac{1}{T}
\sum_{\{ \sigma \}{\in}I^{'}}
{\cal H}(\{\sigma\})
{\rho}_{12}(\sigma_{12})
{\cdots}{\rho}_{kl}(\sigma_{kl})
{\cdots}{\rho}_{N-1N}(\sigma_{N-1N})}
\] 
appearing in both 
numerator and denominator 
of the $\rho_{ij}(\sigma_{ij})$ 
were canceled. 
As the results, 
we obtain $m_{ij}$ as follows. 
\begin{eqnarray}
m_{ij} & = & \sum_{\sigma_{ij}=1}^{Q}
{\sigma}_{ij}{\rho}_{ij}(\sigma_{ij}) \nonumber \\
\mbox{} & = & 
\frac{\sum_{\sigma_{ij}=1}^{Q}{\sigma}_{ij}{\rm e}^{\frac{1}{T}
[J(m_{i,j+1}+m_{i,j-1}+m_{i+1,j}+m_{i-1,j})
+2h\tau_{ij}]\sigma_{ij}-\frac{(2J+h)}{T}
(\sigma_{ij})^{2}}}
{\sum_{\sigma_{ij}=1}^{Q}
{\rm e}^{\frac{1}{T}[
J(m_{i,j+1}+m_{i,j-1}+m_{i+1,j}+m_{i-1,j})
+2h\tau_{ij}]\sigma_{ij}-\frac{(2J+h)}{T}
(\sigma_{ij})^{2}}}.
\end{eqnarray}
If we set $T=1$, we can obtain the 
recursion relations with 
respect to 
the local magnetization  $m_{ij}$ 
under the periodic boundary condition as 
\begin{eqnarray}
m_{ij}^{(t+1)}=
\frac{\sum_{\sigma=1}^{Q}
\sigma {\rm e}^{\omega_{ij}^{(t)}(\sigma)}}
{\sum_{\sigma=1}^{Q}
{\rm e}^{\omega_{ij}^{(t)}(\sigma)}},
\end{eqnarray}
\begin{eqnarray}
{\omega}_{ij}^{(t)}(\sigma)=
\left\{
J[m_{i,j+1}^{(t)}+m_{i,j-1}^{(t)}
+m_{i+1,j}^{(t)}+m_{i-1,j}^{(t)}]
+2h\tau_{ij}\right\}
{\sigma}-(2J+h)\sigma^{2},
\end{eqnarray}
\begin{eqnarray}
m_{i,j+N}=m_{ij},\,\,\,\,\,\,\,\,\,m_{i+N,j}=m_{ij},
\end{eqnarray}
which were 
obtained in the 
previous section as Eqs. (\ref{recur1}), (\ref{recur2}) 
and (\ref{recur3}).

\begin{figure}
\caption{The magnetization of the original image for the case of 
$Q=3$ {\sf (a)} and $Q=4$ {\sf (b)}. The solid lines  
correspond to globally stable solutions. }
\label{fig1} 
\end{figure}

\begin{figure}
\caption{The Hamming distances without 
glassy term ($\beta_{J}=0$) for the case of 
$Q=3$ {\sf (a)}{\sf (d)}. 
The figure {\sf (d)} is 
obtained by expanding the figure {\sf (a)} 
around its minimum. 
The magnetization $m$ and 
corresponding free energy $-f_{\rm RS}$ 
are 
plotted in {\sf (b)} and {\sf (c)}, 
respectively.
Here, the dots lines and the solid lines 
are locally stable states and globally 
stable states, respectively.
} 
\label{static_Q3}
\end{figure}

\begin{figure}
\caption{The Hamming distances without glassy term 
($\beta_{J}=0$) for the case of 
$Q=4$ {\sf (a)}{\sf (d)}. The figure {\sf (d)} is 
obtained by expanding the figure {\sf (a)} 
around its minimum. The magnetization $m$ and corresponding free energy $-f_{\rm RS}$ 
are plotted in {\sf (b)} and {\sf (c)}, respectively.
}
\label{static_Q4}
\end{figure}

\begin{figure}
\caption{
The Hamming distance $D_{\rm H}$ of the MAP 
estimate for the case of $Q=3, T_{s}=0.75, 
\tau_{0}=\tau=1.0$ {\sf (a)}. 
$D_{\rm H}$ is plotted as a function of 
the scaled field $H=h/\beta_{d}$. 
We see that the minimum of $D_{\rm H}$ is 
appeared at $H=H_{\rm opt}=\tau_{0}/2\tau^{2}\beta_{s}=0.375$.
The Hamming distance of the MPM estimate is plotted in {\sf (b)} 
as a function of the temperature $T_{d}$ 
for several values of $H$. 
The figure shows that the minimum of the 
MPM estimate with $H=H_{\rm opt}$ is lower than 
that of the MAP estimate with $H=H_{\rm opt}$. 
}
\end{figure} 
\begin{figure}
\caption{
The Hamming distance $D_{\rm H}$ as a function of the 
strength of the glassy term $\beta_{J}$ 
for several values of $J_{0}$. 
We set $J=1.0$. 
The point $\beta_{J}=0$ 
corresponds to the 
minimum of the Hamming distance in FIG. 2.
}
\label{glass}
\end{figure}
\begin{figure}
\caption{
The time evolutions of the 
Hamming distance are 
plotted in {\sf (a)} 
$T_{d}=T_{s}=0.75$ and {\sf (b)} $T_{d}=0.2\,{\neq}\,T_{s}=0.75$. 
We see that 
for the case of $T_{d}=T_{s}$, 
the Hamming distance converges to its optimal 
value for any initial state of the dynamics.
On the other hand, 
if we set $T_{d}=0.2\,{\neq}\,T_{s}$, 
the Hamming distance converges to the wrong 
state which is higher than the Hamming 
distance between the original image and 
the corrupted one, that is, $D_{\rm H}^{(1)}=0.5$.
}
\label{time_ev}
\end{figure}

\begin{figure}
\caption{
The inverse relaxation time $1/t_{0}$ as 
a function of $T_{d}$ for the case of 
$Q=3, T_{s}=0.75, \tau=\tau_{0}=1.0$.
}
\end{figure}

\begin{figure}
\caption{The Hamming distance 
calculated by Monte Carlo simulation for 
$100 \times 100$ standard picture ``house''. 
The curves averaged over $20$ MCS 
runs are shown  in 
($\beta_{J}=0$ {\sf (a)} and 
$\beta_{J}\,\neq\, 0$ {\sf (b)}).
\label{no_glass}
}
\end{figure}

\begin{figure}
\caption{The original picture {\sf (a)}
(``house'', size $100 \times 100$), the 
corrupted picture by $\sigma=1.2$ Gaussian noise 
{\sf (b)}, the restored pictures at  
$\beta_{J}=0$ {\sf (c)} and $\beta_{J}\neq 0$ {\sf (d)} are displayed}
\label{montecarlo_fig}
\end{figure}

\begin{figure}
\caption{The results of the iterative algorithm 
are displayed. 
The original ``girl'' picture of $128{\times}128$ {\sf (a)}, 
the degraded picture {\sf (b)}, 
the restored pictures with $J=0.2$ {\sf (c)}, 
$J=1.8$ {\sf (d)} and $J=2.5$ {\sf (e)} 
are shown. }
\end{figure}

\begin{figure}
\caption{The results of the iterative algorithm 
are displayed. 
The original ``chair'' picture of $128{\times}128$ {\sf (a)}, 
the degraded picture {\sf (b)}, 
the restored pictures with $J=0.2$ {\sf (c)}, 
$J=1.8$ {\sf (d)} and $J=2.5$ {\sf (e)} 
are shown. }
\end{figure}

\begin{figure}
\caption{The results of the iterative algorithm 
are displayed. 
The original ``house'' picture of $128{\times}128$ {\sf (a)}, 
the degraded picture {\sf (b)}, 
the restored pictures with $J=0.2$ {\sf (c)}, 
$J=1.8$ {\sf (d)} and $J=2.5$ {\sf (e)} 
are shown. }
\end{figure}

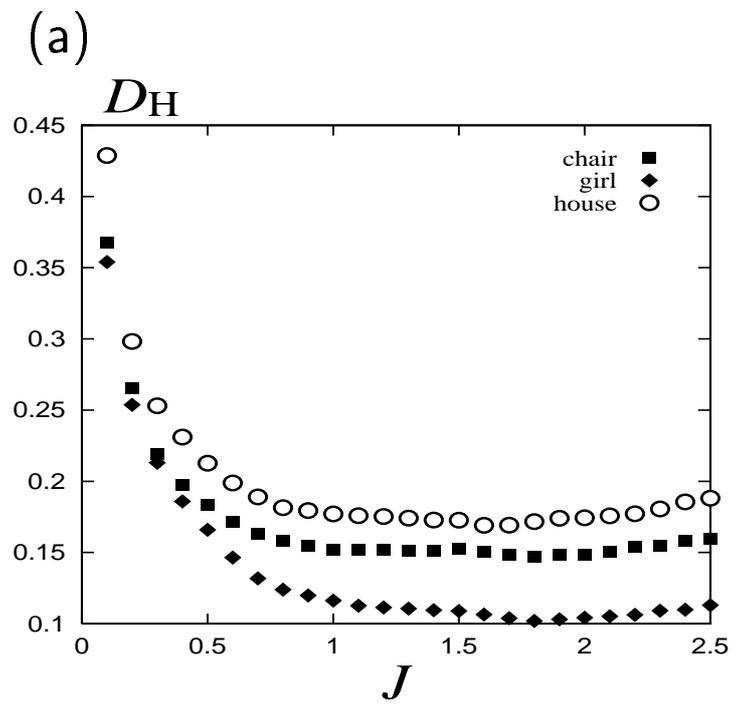
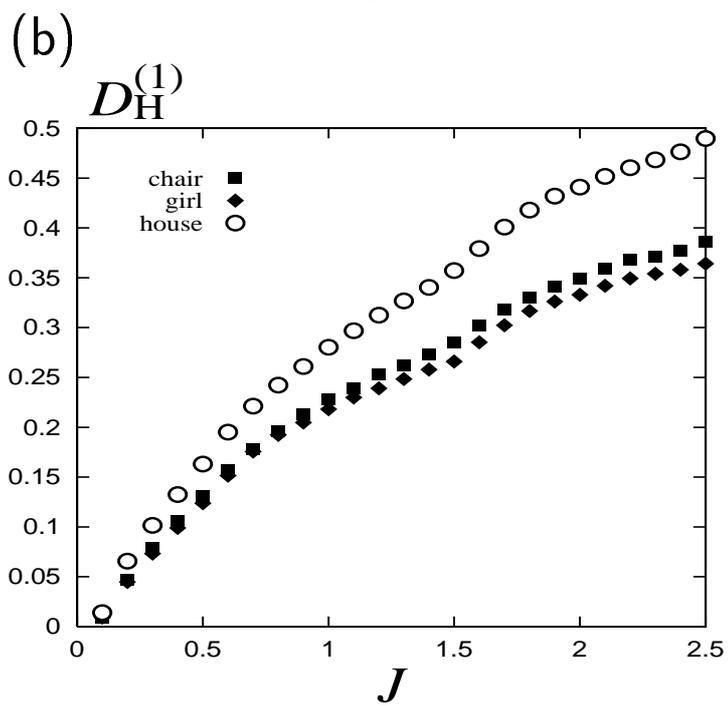
\begin{figure}
\caption{The Hamming distance between 
the original image and the restored one 
$D_{\rm H}$ as a function of $J$ [{\sf (a)}] 
obtained by the iterative algorithm. 
The Hamming distance between 
the restored image and the degraded one 
$D_{\rm H}^{(1)}$ is shown in 
{\sf (b)}.
}
\end{figure}

\begin{thebibliography}{9999}
\bibitem{Marroquin}
J. Marroquin, S. Mitter and T. Poggio, 
J. Am.  Stat.  Assoc. {\bf 82},  76 (1987). 
\bibitem{PB}
J. M. Pryce and A. D. Bruce, 
J. Phys.  A: Math. Gen. {\bf 28},  511 (1995). 
\bibitem{GG}
D. Geiger and F. Girosi, 
IEEE Trans. on Pattern. Anal. Mach. Intel.  {\bf 15},  401 (1991).
\bibitem{NW}
H. Nishimori and K. Y. M. Wong, 
Phys.  Rev.  E {\bf 60},  132 (1999). 
\bibitem{CI}
D. M. Carlucci and J. Inoue, 
Phys.  Rev.  E  {\bf 60},  2547 (1999).
\bibitem{SK}
D. Sherrington and S. Kirkpatrick, Phys. Rev. Lett. {\bf 35}, 1792 (1975).
\bibitem{Bolle}
D. Bolle', H. Rieger and G. M. Shim, 
J. Phys. A : Math. Gen. {\bf 27},  3411 (1994).
\bibitem{Tanaka}
K. Tanaka and T. Morita, Phys. Lett. A {\bf 203}, 122 (1995).
\bibitem{Morita}
T. Morita and K. Tanaka, Physica A {\bf 223}, 245 (1995).
\bibitem{SC}
N. S. Skantzos and A. C. C. Coolen, 
J. Phys. A : Math. Gen. {\bf 33}, 1841 (2000).
\end{thebibliography}
\end{document}